\documentclass[aps,prl,preprint,showpacs,showkeys,groupedaddress]{revtex4}
\newcommand{\W}{14cm}
\newcommand{\Wh}{8cm}
\usepackage{graphicx}
\usepackage{amsmath}
\begin{document}
\title{Hysteresis, force oscillations and non-equilibrium  effects 
in the adhesion of spherical nanoparticles to atomically smooth surfaces}
\author{German Drazer$^\dag$}
\email{drazer@mailaps.org}
\author{Boris Khusid$^\S$}
\author{Joel Koplik $^\ddag$}
\author{Andreas Acrivos $^\dag$}
\affiliation{$^\dag$ Benjamin Levich Institute,
City College of the City University of New York, New York, NY 10031}
\affiliation{$^\S$ Department of Mechanical Engineering, New Jersey Institute of 
Technology, University Heights, Newark, NJ 07102}
\affiliation{$^\ddag$ Benjamin Levich Institute and Department of Physics,
City College of the City University of New York, New York, NY 10031}
\date{\today}
\begin{abstract}
Molecular dynamics simulations are used to examine hysteretic effects and distinctions 
between equilibrium and non-equilibrium aspects of particle adsorption on the walls of 
nano-sized fluid-filled channels.  The force on the particle and the system's Helmholtz 
free energy are found to depend on the particle's history as well as on its radial 
position and the wetting properties of the fluid, even when the particle's motion occurs 
on time scales much longer than the spontaneous adsorption time. The hysteresis is 
associated with changes in the fluid density in the gap between the particle and the wall, 
and these structural rearrangements persist over surprisingly long times. The force and free 
energy exhibit large oscillations with distance when the lattice of the structured nanoparticle 
is held in register with that of the tube wall, but not if the particle is allowed to rotate 
freely.  Adsorbed particles are trapped in free energy minima in equilibrium, but if the 
particle is forced along the channel the resulting stick-slip motion alters the fluid 
structure and allows the particle to desorb.
\end{abstract}
\maketitle

The behavior of fluids in confined geometries with characteristic
length scales in the nanometer range has received enormous attention
in recent years.  Among the motivations for such {\it nanofluidic}
studies is the development of controlled fabrication techniques for
materials at this scale, and their potential application to
``lab-on-a-chip'' devices \cite{StoneSA04} which could perform
analyses of biochemical species at the single-molecule level
\cite{HongQ03}.  Obviously, understanding the transport of suspended
nanometer-size particles under microscopic confinement is a crucial
ingredient in achieving the potential benefits of integrated
nanofluidic systems.  Atomistic numerical simulations have provided
invaluable insight into the behavior of fluids under nanoconfinement,
and have illustrated some dramatic phenomena not present in
macroscopic hydrodynamics \cite{BecksteinS03}.  In
previous work we used molecular dynamics (MD) simulations
to study the transport of closely-fitting nanometer-size particles
driven through a fluid-filled nanochannel, and observed a sharp
transition between steady transport and spontaneous adsorption of
the suspended particles, as a function of the solid-fluid interaction
strength \cite{DrazerKAK02,DrazerKKA05}.  Even for the
computationally-efficient Lennard-Jones interactions used in that
work, we were severely limited in the time scales accessible to
simulation to values $\lesssim 0.1$ ${\mu}$s, and extrapolation to
operational, laboratory time scales is somewhat uncertain.  It
would be very desirable therefore to develop multi-scale simulation
techniques to bridge this gap, and to this end we study here the
same adhesion phenomena from a thermodynamic perspective, exploring
the free energy and fluid configurational changes associated with
the adsorption of nanoparticles.   The free energy is particularly 
relevant to mesoscale calculational methods such as phase field models 
and dissipative particle dynamics. 

In fact, we observe substantial hysteresis
in the Helmholtz free energy of the system when a nanoparticle
executes a controlled adsorption/desorption cycle, even over times
much larger than the time scale of the natural phenomena.
The underlying mechanism appears to be the long-lived structural
rearrangements of the fluid molecules in the gap region between the
particle and the wall, in particular a delay in replenishing fluid in
the gap between the wall and a receding particle.  The persistence of 
this retarded condensation is quite surprising in a
liquid with such simple short-ranged interactions.  Since the free
energy is a state variable, in the thermodynamic limit it should
not exhibit hysteresis, and our results raise qualms about the
conventional assumption of local thermodynamic equilibrium in most
continuum modeling.  In addition to hysteresis, nanoscale particle forces
often exhibit oscillations as a function of radial position, and we observe
that their presence depends on the ability of the particle
to rotate as it approaches the wall.  Oscillations are present when the
underlying lattice of the structured particles has a fixed orientation
relative to that of the tube wall, but not if the particle is 
allowed to rotate freely.  Lastly, we compare the desorption of the particle 
from the wall in equilibrium and non-equilibrium  situations:  in
equilibrium there is a substantial free energy barrier to particle release, whereas
if the particle is driven parallel to the wall it is more likely to detach.

The MD simulations are based on an atomistic description of a fluid
interacting via Lennard-Jones potentials, $V_{LJ}(r)=4\epsilon
\left[(r/\sigma)^{-12}-A (r/\sigma)^{-6}\right]$, where $r$ is the
interatomic separation, $\sigma$ is roughly the size of the repulsive
core, of order a few angstroms, and will be used as the length
scale, $\epsilon$ is the strength of the potential (and the energy
unit), and $A$ is a dimensionless parameter that controls the
attraction between the various atomic species and thus determines
the wetting properties of the system \cite{BarratB99}.  The
corresponding characteristic time scale, $\tau=\sqrt{m \sigma^2 /
\epsilon}$, where $m$ is the mass of the fluid atoms, is a few
picoseconds.  A Nos\'e-Hoover thermostat is used to fix the temperature
of the system at $T=1.0 \epsilon/k_B$, with $k_B$ the Boltzmann
constant.  Further technical details of the simulations may be found
in \cite{DrazerKKA05}.  The fluid is confined to a cylindrical
channel, whose walls are composed of atoms of mass $m_w=100m$,
tethered by a stiff linear spring to fixed lattice sites obtained
from a cylindrical section of an fcc lattice with lattice constant
$\ell=1.71$ corresponding to a wall number density $\rho_w=0.8$.
(Length of the tube: $L_x=34.20$. Inner and outer radii: $R=10.26$
and $R_o=16.25$) The spherical nanoparticle is constructed in an
analogous fashion using a spherical section of radius $a=5.13$. The
atoms in the particle are, however, fixed at the lattice sites,
allowing the motion of the nanoparticle to be computed by rigid
body dynamics, integrating Newton's and Euler's equations.  Finally,
to avoid a perfect matching in the underlying structures of both
solids the equilibrium  position of each atom is perturbed by a
small random displacement \cite{DrazerKKA05}.  The particle and
channel wall are made of the same material and their atoms interact
via the LJ potential with $A=1$.  The fluid atoms also interact
among themselves with the $A=1$ standard LJ potential and the
volume-average number density is the same as that of the tube wall,
$\rho_{av}=0.8$, which, at $T=1.0$, corresponds to the fluid phase
of the Lennard-Jones bulk system, slightly above the critical
temperature.

In our previous work, we applied an axial force to the particle, and
observed a sudden spontaneous transition from statistically
steady motion parallel to the tube axis to adsorption on the walls,
when the strength of the solid-fluid van der Waals attraction falls
below a critical value $A_c\simeq 0.8$ (for both spherical and
ellipsoidal particles).  Here, we first measure the changes in
the Helmholtz free energy $\cal F$ of the system as the sphere 
{\em quasi-statically} approaches the wall in the radial direction 
and then recedes to its original position at the center of the tube,
by performing a thermodynamic integration. The free energy difference between any two
points during this switching process is then given by ${\cal
F}(r)-{\cal F}(0)=W=\int_0^r \langle F(\lambda) \rangle ~d\lambda$,
where $W$ is the work done on the system and $\langle F \rangle$
is the radial force acting on the particle at radius $\lambda$,
averaged over an ensemble of 10 statistically independent realizations,
and over a time interval $t_s$. Specifically, the center of mass
of the sphere is translated at constant velocity in the radial
direction, typically over a distance $\Delta r=0.01$ in time $t_s=50$
($u=2\times10^{-4}\sim1$cm/s), then held in place for a further time
interval $t_s$, then translated again. First, we consider the case where
the sphere is allowed to
rotate freely in the suspending fluid at all times.  The time average
is meant as a substitute for an ensemble average, assuming ergodicity,
with the (modest) ensemble average is meant to compensate for any
peculiarities in the initial configuration.

\begin{figure}[t] 
\includegraphics*[width=\W]{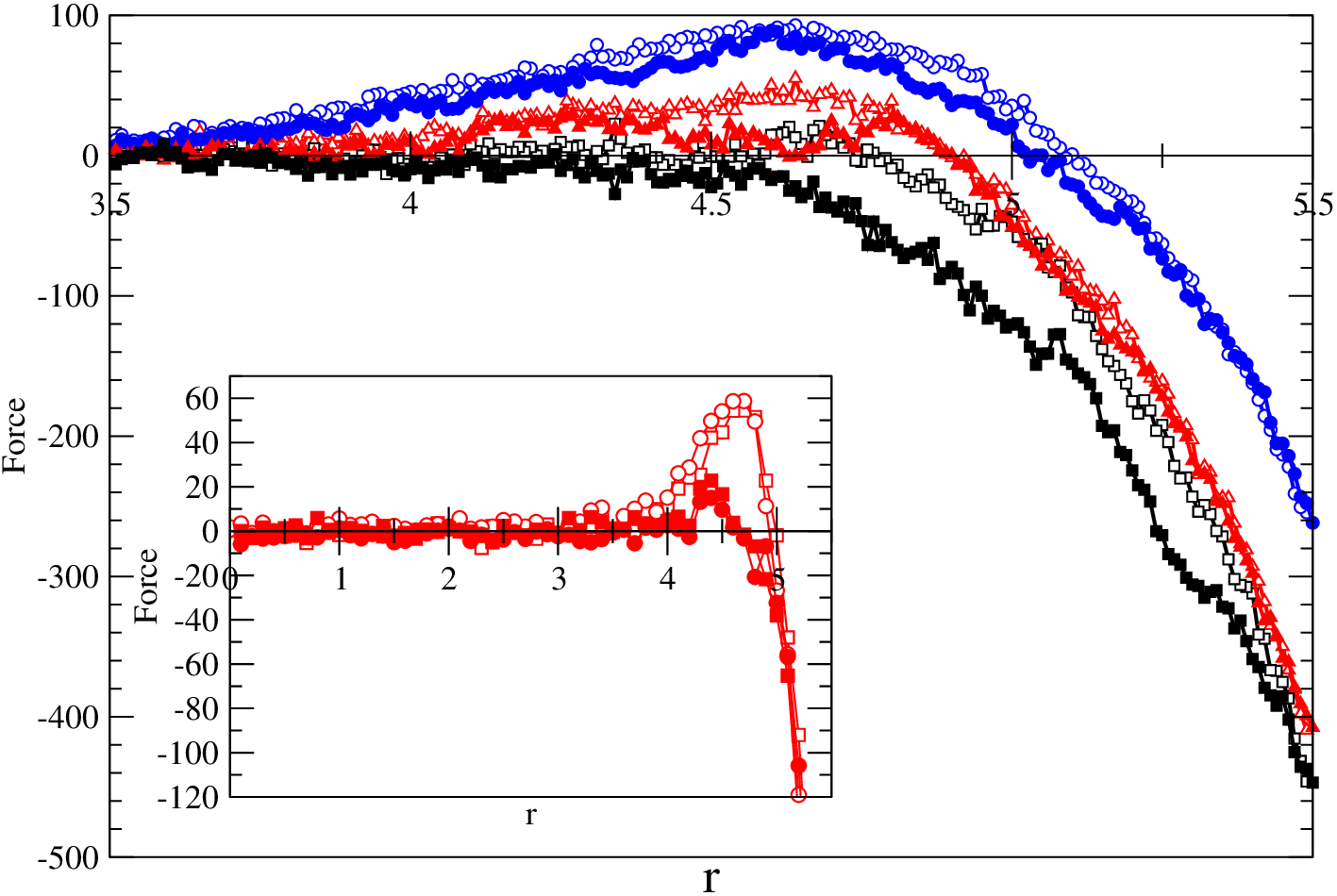} 
\caption{\label{force}
Average force on the spherical particles during the adsorption/desorption
cycles as a function of the radial position for $A=0.6$ (circles),
$A=0.8$ (triangles), and $A=1$ (squares).  The solid (open) symbols
correspond to the approaching (receding) part of the cycle.
The approaching velocity is $u\sim1$cm/s. 
The inset shows, for the case $A=0.8$, the average force
measured during radial motion (squares) and in a stationary radial
position (circles) for an overall approaching speed of $u\sim10$cm/s.}
\end{figure}

\begin{figure} 
\includegraphics*[width=\W]{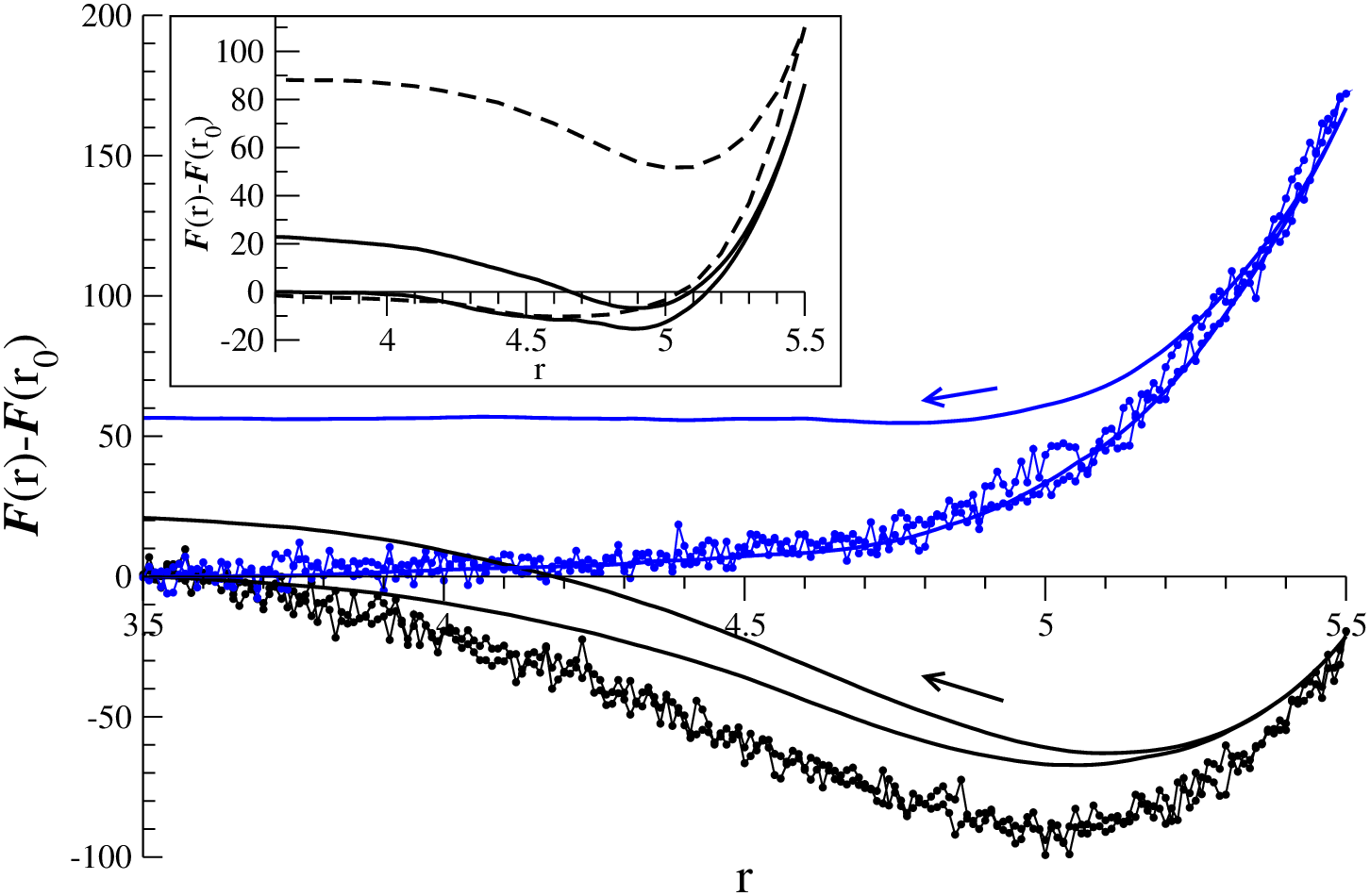} 
\caption{\label{f}
Free-energy and potential-energy differences as a function of the
radial position of the spherical particle for $A=0.6$ (bottom curves)
and $A=1$ (top curves). The points correspond to the potential
energy of the system evaluated from the interparticle interactions.
The solid line corresponds to the free-energy calculated during the
adsorption/desorption cycle (an arrow indicates the curve corresponding
to the receding part of the measurement.) The inset shows the
free-energy differences computed for $A=0.8$ at two different
approaching velocities, $u\sim10$cm/s (dashed lines) and $u\sim1$cm/s
(solid line).} 
\end{figure}

In Fig.~\ref{force} we show the average force on the sphere as a
function of its radial position, for various degrees of wetting,
$A=0.6$, 0.8 and 1.0, which exhibits large hysteresis as $A\to1$.
On the other hand, we found no hysteresis, 
even at relatively large approaching velocities,
for the $A=0$ case of a completely non-wetting fluid.
(An isolated fluid drop with this interaction would float off a solid wall.)
The inset to Fig.~\ref{force} compares, for one case, the
average force during  the translation step to that when the sphere
is held in place, and the agreement between the two values indicates
that the observed hysteresis is not due to drag or other irreversible
forces arising from the motion of the sphere, in spite of the
relatively large approaching speed ($u\sim10$cm/s) used in this 
comparison.  In Fig.~\ref{f} we present the Helmholtz free-energy
of the system, together with the potential energy, for the same values
of $A$.  The adsorption/desorption hysteresis is clear in the $A>0$ 
cases, with the work needed to separate the particles from the wall
clearly higher that the gain in free-energy from bringing the
particles in contact with the wall, {\it i.\ e.}, $W >0$. This result
shows that the adhesion process is generally a non-equilibrium process that
dissipates energy.  On the other hand, we observe that the potential energy of the
system (sum of all LJ interactions) does not display any significant hysteresis. In addition, a
detailed study of each of the separate contributions shows that the energy
increase observed for $r\gtrsim5.0$ in the wetting case $A=1$
results from the deformation of the wall as the particle pushes
against it, rather than structural or solvation effects.

\begin{figure}[th] 
\includegraphics*[width=\Wh]{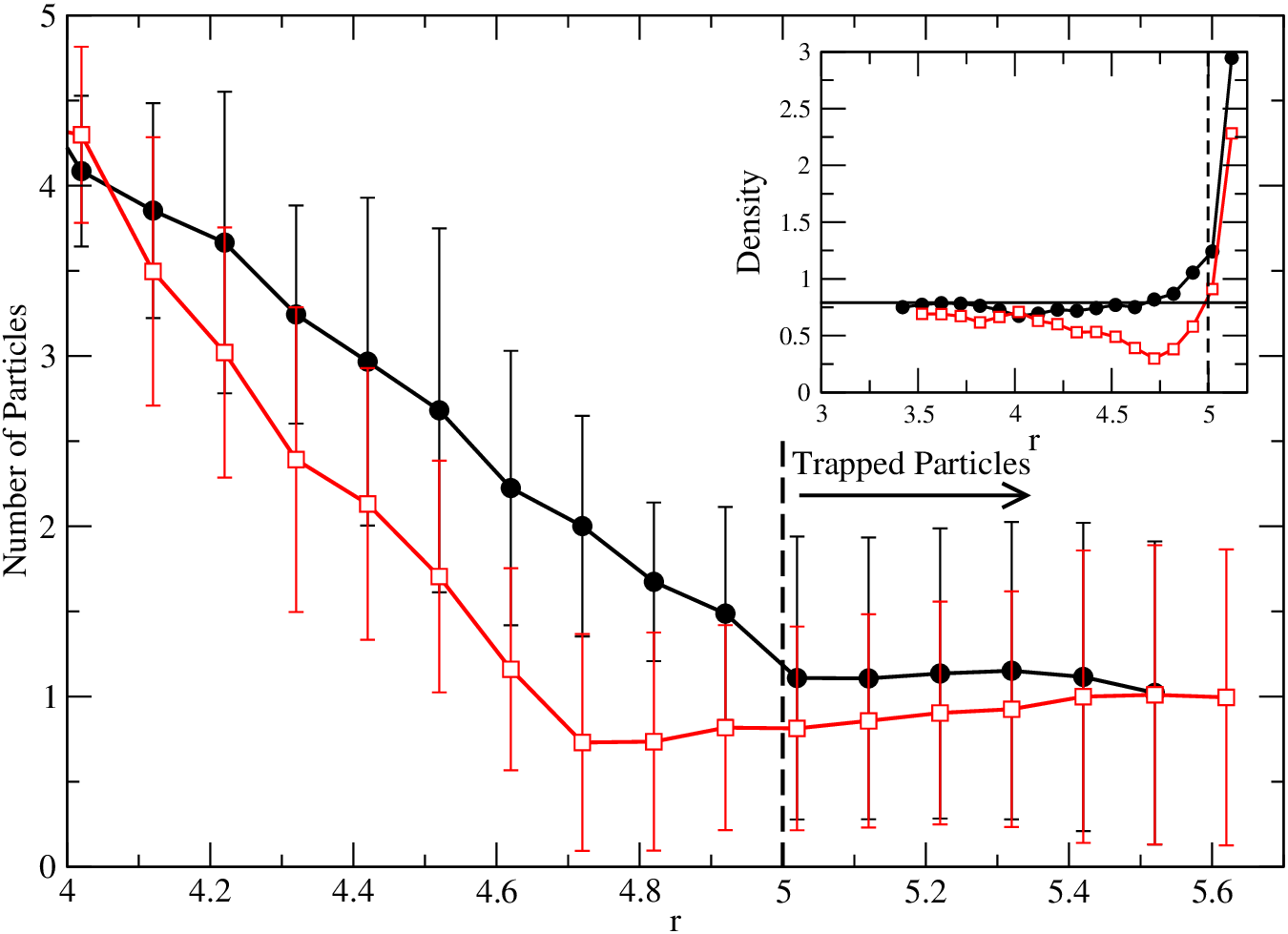}
\includegraphics*[width=\Wh]{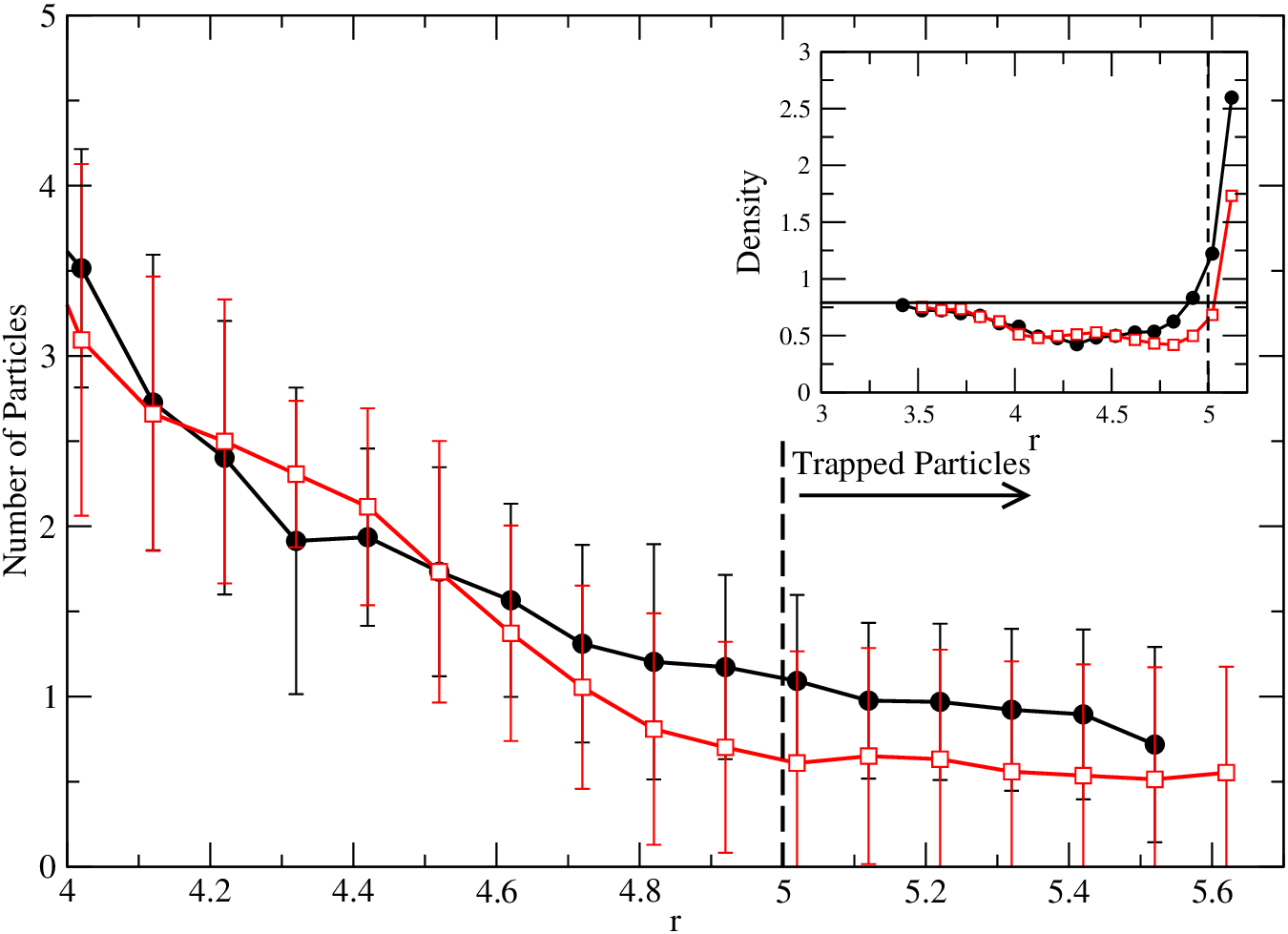} 
\caption{\label{density}
Average number of fluid molecules in the particle-wall gap during
the adsorption/desorption cycles for $A=0.85$. The number of particles 
is tracked in a cylindrical region of radius $a/4$ between the sphere and the tube
surfaces, and the corresponding number density is given in the
insets. The solid (open) symbols correspond to the approaching (receding) branch and
the dashed line marks the radial position at which the volume of the cylindrical
region is less than the volume of a single fluid molecule. 
The top (bottom) figure corresponds to $u\sim10$ cm/s ($u\sim1$cm/s).} 
\end{figure}


\begin{figure}[t] 
\includegraphics*[width=\W]{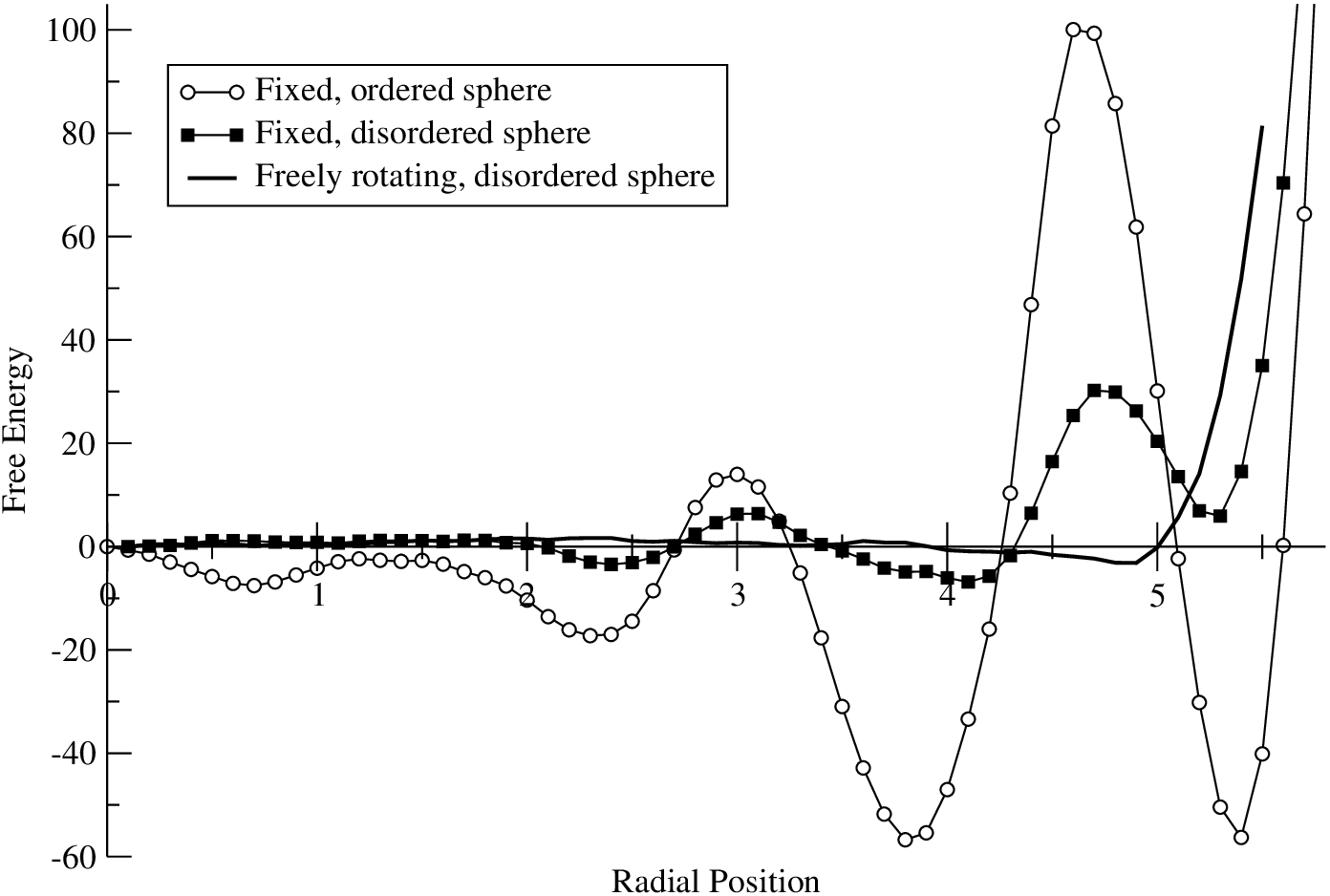}
\caption{\label{orientation}
Free-energy differences as a function of the radial position of the spherical 
particle for $A=1$. The open (closed) symbols correspond to an ordered (disordered) 
sphere with fixed orientation approaching an ordered (disordered) tube wall. 
The solid line corresponds to the disordered sphere rotating freely.} 
\end{figure}

We have performed simulations at different approach speeds and for 
different wetting conditions and, although $W>0$ in all $A>0$ cases, 
the hysteresis is reduced as the characteristic times 
become larger, as shown in the inset to Fig.~\ref{f}.  This is
expected, since the process should be reversible with $W=0$
if it were to be carried out infinitely slowly \cite{Israelachvili92}.  
It is important to note, however, that the motion of the sphere in
these simulations is very slow compared with the natural time scale of the
spontaneous process.  Indeed, the characteristic time for the approaching 
and receding parts of the free energy measurements is roughly 
$2.5\times10^4\tau$, an order of magnitude larger than typical 
adsorption times observed when a sphere is freely suspended. (See, 
for example, the estimate of the diffusive time to reach the wall,
$\tau_D\sim2500\tau$, and the numerical results in Table II for $A=0.6$,
in Ref.~\cite{DrazerKKA05}.)

The molecular mechanism underlying the 
hysteresis appears to be the history-dependence of the number of particles
in the gap between the sphere and the tube wall, for gap sizes of the order 
of the diameter of a single fluid molecule.  In Fig.~\ref{density} we see
that the number of gap particles during the desorption branch is smaller, 
corresponding to a depletion of particles relative to the bulk density.   
The absence of fluid leads to the dominance of the attractive wall-particle 
interactions and depletion forces, explaining the results presented in
Fig.~\ref{force}, including the surprising result obtained
for $A=0.80$ at $u\sim10$cm/s, in which the force on the sphere is
repulsive when approaching the wall but attractive in the pull-off
measurements (see the inset to Fig.~\ref{force}).  
The dependence of hysteresis on the degree of wetting of the fluid may be
understood in terms of the relative ease of pushing fluid atoms out of the gap
region, which obviously improves as the fluid becomes less wetting. 
Furthermore, the presence of a nearly-planar wall induces layering and
presumably other structural correlations within the fluid, which can
contribute to hysteresis if any significant time is required for the 
structure to reestablish itself after the particle is pulled away.  
Again, this effect is reduced as $A\to 0$. Lastly,
we note from Fig.~\ref{density} that the
hysteresis in the number of gap particles decreases substantially 
as the approach velocity is reduced, consistent with the presumption that it
would be absent in the limit of infinitesimally slow approach.

Somewhat analogous hysteretic behavior due to fluid depletion has been seen 
in a study of cavitation by Bolhuis and Chandler \cite{BolhuisC00}, who
measured the force between non-wetting parallel plates which approach and
recede while immersed in a Lennard-Jones liquid.  These authors consider a
WCA potential, whose behavior one would expect to be similar to our 
$A=0$ case, but other operating conditions differ in that
the bulk density and temperature are much lower, while the plate velocity is
much higher.  In rough agreement with our results, 
force hysteresis is not observed at the highest temperature
studied, $T=0.81\epsilon/k_B$, but does appear at lower temperatures,
perhaps as a result of glassy behavior.  In the latter cases, there are high
and low density states in the gap between plates with the high
density state apparently metastable, and the hysteresis is attributed 
to a retardation in the (non)wetting-induced cavitation inside the gap.

A notable feature of the density profiles in Fig.~\ref{density} is 
that the particles leave the gap {\it continuously}, without a noticeable 
effect on the solvation layers known to occur close to solid surfaces
\cite{Israelachvili92,McGuigganI88,QinF03}. The smoothness of this transition 
is in agreement with the absence of oscillatory solvation forces in the 
measurements presented in Fig.~\ref{force}.  This effect results from the
ability of the particle to rotate freely as it approaches the wall,
and indicates that a freely suspended particle might overcome solvation force
barriers that prevent strong adhesion, by adjusting its atomic lattice orientation
relative to that of the wall.
In Fig.~\ref{orientation} we show the results of an alternate simulation at $A=1$ 
where the
orientation of the particle is fixed during its slow motion towards the
wall, which does exhibit the familiar oscillations as layers of fluid are
successively squeezed out of the gap region.  
In contrast, in the freely rotating case the structured particles have some freedom to avoid close 
approach to a particular fluid atom by a rotation of their atomic lattice, leading to a gradual
displacement of fluid atoms, in contrast to the fixed case where the 
approaching particle uniformly displaces fluid away from itself.
Furthermore, when the particle is very close to the wall, its orientation 
tends to lock into values determined by the nearly solid wall structure.
Note that the disorder introduced in the position of the solid atoms
is smaller than the atomic diameter of the
fluid atoms and should not destroy the molecular ordering of the fluid 
close to the wall \cite{Israelachvili92} 
and in fact density oscillations are observed in our system 
\cite{DrazerKKA05}.  The same argument applies to the curvature of the 
tube wall. In fact, we also show in fig.~\ref{orientation} that, 
although substantially reduced compared to the ordered case, oscillatory 
solvation forces are still present for disordered solids.


The particle desorption process also  exhibits some important physical
distinctions depending on whether the particle is left alone in equilibrium 
or is subject to continued forcing, as in \cite{DrazerKAK02,DrazerKKA05}.  
We see in Fig.~\ref{f} that, for partially-wetting fluids ($A<1$),
a particle close to the wall is confined by a substantial potential barrier,
roughly $40\epsilon/k_B$ (for $A=0.8$ and $u\sim1$cm/s), and is unlikely 
to spontaneously desorb under the action of equilibrium fluctuations alone.
However, in our previous study of a driven particle motion we observed 
a spontaneous desorption of particles whenever $A\gtrsim0.8$, which would 
indicate a much smaller pull-off force.  The distinction is that desorption 
only occurs in non-equilibrium simulations in which a
large external force is applied to the particles ($F\sim$pN).  Note however
that the effect is indirect, since the force is not radial but is instead 
in the direction of the tube axis.  As found previously,
axial forcing of the particle 
induces stick-slip motion along the wall after adsorption, which
enhances dramatically
both the mean and rms fluctuation in the number of atoms in the gap region, 
from a value of $2.0\pm0.9$  while in equilibrium to $7.0\pm2.9$ 
when the particle is forced along the wall. Related to this, 
adsorbed spheres undergoing stick-slip motion present 
comparatively large fluctuations in their radial position.
A possible mechanism for desorption is that the sphere encounters ordered 
clusters of fluid molecules as it moves intermittently along the wall, 
which provide an (inward) radial fluctuation to a position where the 
wall attraction is reduced.   

The results in this paper underline some of the issues arising when particle
dynamics at the nanoscale are examined in quantitative detail. 
The results pose some challenges for the long-term goal of 
modifying the continuum description of particle-fluid dynamics to describe 
phenomena at these length scales, since for example there is no single 
unique potential of mean force which may be incorporated.  
One immediate avenue for further study, would be to explore
the validity of the number of particles in the gap as an order parameter for a
multi-scale description.

This work was supported by the Engineering Research Program, office
of Basic Energy Sciences, U. S.  Department of Energy under Grant
No. DE-FG02-03ER46068.

\end{document}